\documentclass{article}
\usepackage{cite}
\usepackage{graphicx}
\usepackage{dcolumn}

\begin{document}

\date{}
\title{Comment on: ``On the characteristic polynomial of an effective
Hamiltonian''}
\author{Francisco M. Fern\'{a}ndez \thanks{%
E-mail: fernande@quimica.unlp.edu.ar} \\
INIFTA, DQT, Blvd. 113 S/N, Sucursal 4, Casilla de Correo 16,\\
1900 La Plata, Argentina.}
\maketitle

\begin{abstract}
We show that a method proposed recently, based on the characteristic
polynomial of an effective Hamiltonian, had been developed several years
earlier by other authors in a clearer and more general way. We outline both
implementations of the approach and compare them by means of the calculation 
of the exceptional point closest to origin for a toy model.
\end{abstract}

In a recent paper published in this journal Zheng\cite{Z22} proposed a
method for improving the convergence properties of the perturbation series.
The approach is based on the characteristic polynomial of an effective
Hamiltonian that should not be obtained explicitly. In principle, the method
is restricted to a finite vector space that can be separated into a $P$%
-space and a $Q$-space and one should focus on the former. Zheng applied the
method to a trivial problem posed by a tridiagonal matrix of dimension $3$.

Apparently, Zheng was unaware that the same approach was put forward several
years earlier by Fried and Ezra\cite{FE89} with somewhat similar arguments.
These authors were interested in nearly resonant molecular vibrations and
applied the approach to the Barbanis Hamiltonian. The resummation method of
Fried and Ezra, based on the reconstruction of the effective secular
equation, is identical to Zheng's one although the former was introduced in
a more general way which does not require that the state space be finite.

In what follows we outline and compare the methods proposed by Zheng\cite%
{Z22} and Fried and Ezra\cite{FE89}.

Zheng's method\cite{Z22} applies to a quantum-mechanical model defined on a
state space of dimension $\tilde{N}=N+M$ spanned by an orthonormal basis set
$B=\left\{ \psi _{i}^{P},\;i=1,2,\ldots ,N,\;\psi _{j}^{Q},\;j=1,2,\ldots
,M\right\} $, where $\left\{ \psi _{i}^{P}\right\} $ and $\left\{ \psi
_{j}^{Q}\right\} $ span the so-called $P$ and $Q$ subspaces, respectively.
The approach is based on the secular determinant
\begin{equation}
\left| E\mathbf{I}_{\tilde{N}}-\mathbf{H}(\lambda )\right| =E^{\tilde{N}%
}+\sum_{j=1}^{\tilde{N}}p_{j}(\lambda )E^{\tilde{N}-j},  \label{eq:sec_det}
\end{equation}
where $\mathbf{H}(\lambda )$ is the matrix representation of the Hamiltonian
operator $H=H_{0}+\lambda H_{I}$ in the basis set $B$ and $\mathbf{I}_{%
\tilde{N}}$ is the identity matrix of dimension $\tilde{N}$. The operators $%
H_{0}$ and $H_{I}$ are the diagonal and off-diagonal parts of $H$ in the
basis set $B$. It is clear that $p_{j}(\lambda )$ are polynomial functions
of $\lambda $.

Zheng\cite{Z22} took into account that
\begin{equation}
\left| E\mathbf{I}_{\tilde{N}}-\mathbf{H}(\lambda )\right| =\prod_{i=1}^{N}%
\left[ E-E_{i}^{P}(\lambda )\right] \prod_{j=1}^{M}\left[ E-E_{j}^{Q}(%
\lambda )\right] ,  \label{eq:sec_det_wrong}
\end{equation}
where $E_{i}^{P}(\lambda )$ and $E_{j}^{Q}(\lambda )$ are determined by the
conditions $E_{i}^{P}(\lambda =0)=\left\langle \psi _{i}^{P}\right| H\left|
\psi _{i}^{P}\right\rangle $ and $E_{j}^{Q}(\lambda =0)=\left\langle \psi
_{j}^{Q}\right| H\left| \psi _{j}^{Q}\right\rangle $ . It is worth noting
that any eigenfunction $\psi _{k}$ of $H$ is a linear combination of both $%
\psi _{i}^{P}$ and $\psi _{j}^{Q}$ because $\left\langle \psi
_{i}^{P}\right| H_{I}\left| \psi _{j}^{Q}\right\rangle \neq 0$.

Zheng's approach is based on the equation
\begin{equation}
\left| E\mathbf{I}_{N}-\mathbf{H}_{\mathrm{eff}}(\lambda )\right|
=\prod_{i=1}^{N}\left[ E-E_{i}^{P}(\lambda )\right] ,
\label{eq:sec_det_Zheng_app}
\end{equation}
where each $E_{i}^{P}(\lambda )$ is a perturbation expansion
\begin{equation}
E_{i}^{P}(\lambda )=\sum_{j=1}^{K}E_{i,j}^{P}\lambda ^{j}.
\label{eq:E_n^P_PT}
\end{equation}
Since the calculation is based on the right-hand-side of equation (\ref%
{eq:sec_det_Zheng_app}) it is not necessary to find out the effective
Hamiltonian operator $H_{\mathrm{eff}}$ explicitly.

Zheng\cite{Z22} applied the approach just outlined to a simple model with $%
N=2$ and $M=1$ and showed that the approximate eigenvalues obtained in this
way are considerably more accurate than the perturbation series used in the
construction of the effective characteristic polynomial (\ref%
{eq:sec_det_Zheng_app}). However, we may raise the following question: why
would anybody resort to this approach since the calculation of the exact
eigenvalues as roots of the polynomial (\ref{eq:sec_det}) is
straightforward?.

As stated above, Fried an Ezra\cite{FE89} put forward essentially the same
method in a more general way by means of somewhat similar arguments. Suppose
that we want to obtain the solutions to the Schr\"{o}dinger equation $H\psi
_{n}=E_{n}\psi _{n}$, $n=1,2,\ldots $, where $H=H_{0}+\lambda H_{I}$. To
this end, we apply perturbation theory and obtain partial sums of the form
\begin{equation}
E_{n}^{[K]}=\sum_{j=0}^{K}E_{n,j}\lambda ^{j}.  \label{eq:E_n^[K]}
\end{equation}%
With them one reconstructs the effective secular equation
\begin{equation}
\left\{ \prod_{n=1}^{N}\left[ W-E_{n}^{[K]}(\lambda )\right] \right\}
^{[K]}=W^{N}+\sum_{j=1}^{N}p_{j}(\lambda )W^{N-j},  \label{eq:F(W,lamb)}
\end{equation}%
where $\left\{ ...\right\} ^{[K]}$ means that we remove any term with $%
\lambda ^{j}$ if $j>K$. The reason is that the accuracy of the results
cannot be greater than $\mathcal{O}\left( \lambda ^{K}\right) $ determined
by the partial sums (\ref{eq:E_n^[K]})\cite{FE89}. The eigenvalues $E_{n}$
used in this reconstruction are related to the so-called model space, which
is finite, and we leave aside the complement space that is not necessarily
so. The states in the model space are strongly coupled among themselves and
weakly coupled to the states in the complement state.

Zheng\cite{Z22} applied the method to the following tridiagonal matrix of
dimension $\tilde{N}=3$:
\begin{equation}
\mathbf{H}(\lambda )=\left(
\begin{array}{lll}
2 & \lambda & 0 \\
\lambda & 1.1 & \lambda \\
0 & \lambda & 1%
\end{array}
\right) .  \label{eq:H_3x3}
\end{equation}
From the roots of the characteristic polynomial
\begin{equation}
cp(E,\lambda )=\left| E\mathbf{I}_{3}-\mathbf{H}\right| =E^{3}-\frac{41}{10}%
E^{2}+\frac{53-20\lambda ^{2}}{10}E+\frac{15\lambda ^{2}-11}{5},
\label{eq:charpoly_N=3}
\end{equation}
we obtain three eigenvalues $E_{1}(\lambda )<E_{2}(\lambda )<E_{3}(\lambda )$
that satisfy $E_{1}(0)=1$, $E_{2}(0)=1.1$ and $E_{3}(0)=2$. We do not show
the exact analytical expressions for $E_{i}(\lambda )$ because they are
rather cumbersome. These eigenvalues do not cross because $H_{12}=H_{21}\neq
0$ and $H_{23}=H_{32}\neq 0$ (see\cite{F14} for a general analysis). Note
that $E_{1}(0)$ and $E_{2}(0)$ are nearly degenerate, a fact that commonly
leads to perturbation series with poor convergence properties (as discussed
by Fried and Ezra\cite{FE89} and Zheng\cite{Z22}).

From the discriminant of the characteristic polynomial (\ref{eq:charpoly_N=3}%
) (see\cite{AF21} and references therein) we obtain the exceptional points $%
\lambda _{1}$, $\lambda _{1}^{*}$, $\lambda _{2}$, $\lambda _{2}^{*}$, $%
-\lambda _{2}$ and $-\lambda _{2}^{*}$, where $\lambda _{1}=0.05139217757i$
and $\lambda _{2}=0.2381164319+0.5028706167i$. This particular distribution
of the exceptional points in the complex $\lambda $-plane is due to the fact
that $cp(E,-\lambda )=cp(E,\lambda )$ and $cp(E,\lambda )^{*}=cp(E,\lambda )$
for $\lambda $ real. The eigenvalues $E_{1}$ and $E_{2}$ coalesce at $%
\lambda =\pm \lambda _{1}$ while $E_{2}$ and $E_{3}$ coalesce at $\lambda
_{2}$ and its variants. Therefore, following the arguments put forward by
Fried and Ezra\cite{FE89} and Zheng\cite{Z22}, we apply the approach (\ref%
{eq:F(W,lamb)}) for $N=2$ (that is to say: with perturbation approximations (%
\ref{eq:E_n^[K]}) for $E_{1}$ and $E_{2}$).

\begin{figure}[tbp]
\begin{center}
\includegraphics[width=9cm]{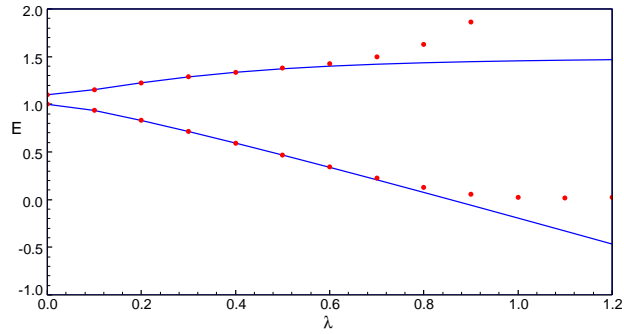}
\end{center}
\caption{First two eigenvalues $E_1(\protect\lambda)$ and $E_2(\protect%
\lambda)$ calculated exactly (blue, continuous line) and by means of the
approach (\protect\ref{eq:F(W,lamb)}) based on sixth-order perturbation
theory}
\label{Fig:PT6}
\end{figure}

Since $cp(E,-\lambda )=cp(E,\lambda )$ we have perturbation expansions in
terms of the variable $\lambda ^{2}$. Figure~\ref{Fig:PT6} shows results
obtained from equation (\ref{eq:F(W,lamb)}) with perturbation expansions of
order $K=6$. These results are similar to those shown by Zheng\cite{Z22}.

A most interesting feature of this approximation is that it enables us to
estimate the exceptional point in the complex $\lambda $ plane closest to
origin ($\lambda _{1}$ and $\lambda _{1}^{\ast }$ in the present case). This
fact was overlooked by Fried and Ezra because the perturbation series for
the eigenvalues of the Barbanis Hamiltonian are divergent (zero convergence
radius). On the other hand, Zheng estimated the location of the exceptional
point by means of the approach, using perturbation series of order $K=2,4,6$. 
In order to illustrate this fact explicitly, we apply the discriminant
approach\cite{AF21} to equation (\ref{eq:F(W,lamb)}) and obtain the
approximate exceptional points for increasing values of $K$. Table ~\ref%
{tab:Ep} shows that the approximate exceptional point $|\lambda _{1}|^{[K]}$
calculated in this way converges, as $K$ increases, towards the actual value
of $|\lambda _{1}|$ obtained from the characteristic polynomial (\ref%
{eq:charpoly_N=3}). Zheng's result for $K=2$ appears to be incorrect because
it agrees with the result for $K=6$ and disagrees with present one. The
agreement of present results with those of Zheng for $K=4,6$ clearly reveal
that the methods of Fried and Ezra and Zheng are identical; the discrepancy
for $K=2$ being, most probably, due to a misprint in Zheng's paper.

\textit{Summarizing}: the aim of this Comment is to point out that Zheng\cite%
{Z22} did not develop a new method because it was put forward several years
earlier by Fried and Ezra\cite{FE89}. Besides, the latter authors presented
their approach in a clearer and more general way. In addition to it Fried
and Ezra applied the reconstruction of the effective secular equation to
more demanding quantum-mechanical models. As a by-product we verify that the
method is suitable for estimating the locations of the exceptional points
that determine the radius of convergence of the perturbation series used in
the reconstruction of the effective secular equation.

\begin{table}[tbp]
\caption{Convergence of the estimated exceptional point $|\protect\lambda%
_1^{[K]}|$}
\label{tab:Ep}
\begin{center}
\begin{tabular}{rll}
\hline
$K$ & \multicolumn{1}{c}{Present} & \multicolumn{1}{c}{Zheng\cite{Z22}} \\ \hline
2 & 0.05147186257 & 0.0513922 \\
4 & 0.05139244862 & 0.0513924 \\
6 & 0.05139217790 & 0.0513922 \\
8 & 0.05139217757 &  \\
10 & 0.05139217757 &
\end{tabular}%
\end{center}
\end{table}

\begin{thebibliography}{9}
\bibitem{Z22} Y. Zheng, Phys. Lett. A 443 (2022) 128215.

\bibitem{FE89} L. E. Fried and G. S. Ezra, J. Chem. Phys. 90 (1989)
6378-6390.

\bibitem{F14} F. M. Fern\'{a}ndez, J. Math. Chem. 52 (2014) 2322.

\bibitem{AF21} P. Amore and F. M. Fern\'andez, Eur. Phys. J. Plus 136 (2021)
133.
\end{thebibliography}
\end{document}